\documentclass[aps,prb,twocolumn,showpacs,superscriptaddress,preprintnumbers,10pt]{revtex4-1}
\usepackage{graphicx} %Include figure files,superscriptaddress
\usepackage{amsmath}

\newcommand{\be}{\begin{equation}}
\newcommand{\ee}{\end{equation}}
\newcommand{\bea}{\begin{eqnarray}}
\newcommand{\eea}{\end{eqnarray}}
\newcommand{\bse}{\begin{subequations}}
\newcommand{\ese}{\end{subequations}}

\begin{document}

\title{Multiple magnetic transitions in the spin-$\frac12$ chain antiferromagnet SrCuTe$_{2}$O$_{6}$}

\author{N. Ahmed}
\affiliation{School of Physics, Indian Institute of Science
Education and Research Thiruvananthapuram-695016, India}
\author{A. A. Tsirlin}
\email{altsirlin@gmail.com}
\affiliation{National Institute of Chemical Physics and Biophysics, 12618 Tallinn, Estonia}
\affiliation{Experimental Physics VI, Center for Electronic Correlations and Magnetism, Institute of Physics, University of Augsburg, 86135 Augsburg, Germany}
\author{R. Nath}
\email{rnath@iisertvm.ac.in}
\affiliation{School of Physics, Indian Institute of Science
Education and Research Thiruvananthapuram-695016, India}
\date{\today}

\begin{abstract}
Using thermodynamic measurements and density-functional band-structure calculations, we explore magnetic behavior of SrCuTe$_2$O$_6$. Despite being a structural sibling of a three-dimensional frustrated system PbCuTe$_2$O$_6$, this spin-$\frac12$ quantum magnet shows remarkably different low-temperature behavior. Above 7\,K, magnetic susceptibility of SrCuTe$_2$O$_6$ follows the spin-chain model with the antiferromagnetic intrachain coupling of $J\simeq 49.3$\,K. We ascribe this quasi-one-dimensional behavior to the leading third-neighbor coupling that involves a weakly bent \mbox{Cu--O$\ldots$O--Cu} superexchange pathway with a short O$\ldots$O contact of 2.79\,\r A. Below 5\,K, SrCuTe$_2$O$_6$ undergoes two consecutive magnetic transitions that may be triggered by the frustrated nature of interchain couplings. Field dependence of the magnetic transitions (phase diagram) is reported.
\end{abstract}
\maketitle

\section{Introduction}
Geometrically frustrated magnets are difficult to construct.\cite{Ramirez1994,Greedan2001} Their exotic properties rely on the fact that antiferromagnetic (AFM) couplings on a triangular loop are equal, as in the ideal kagome, hyperkagome, or pyrochlore spin lattices.\cite{Moessner2006,Balents2010} Geometrical distortions render magnetic couplings nonequivalent, thus alleviating the frustration, lifting the classical degeneracy and eventually stabilizing conventional ordered ground states. Materials featuring high crystallography symmetry and triangular-like structural features are indispensable for the field of frustrated magnetism, because multiple symmetry elements of the crystal structure ensure that the couplings on different bonds of the triangular loop are equal, and the strong magnetic frustration persists. Real-world examples of such systems are particularly rare for the case of \mbox{spin-$\frac12$},\cite{Mendels2010,Okamoto2007,Dally2014} where strongest quantum effects and, hence, the ultimate elimination of conventional ordered ground states is expected.

Recently, Koteswararao \textit{et al}.\cite{Koteswararao2014} reported synthesis and magnetic behavior of PbCuTe$_2$O$_6$. This cubic material entails a hyperkagome network of Cu$^{2+}$ spins connected by the second-neighbor AFM coupling. However, first-neighbor couplings forming isolated triangles and third-neighbor couplings forming spin chains are present as well, thus rendering the spin lattice highly complex. Experimental results put forward an overall AFM behavior with the Curie-Weiss temperature of 22\,K, yet neither a broad maximum of the susceptibility nor a long-range ordering transition are seen above 1.2\,K. Around 0.87\,K, a kink in the heat capacity indicates a magnetic transition of unknown origin.

Here, we explore magnetic behavior of SrCuTe$_2$O$_6$,\cite{Wulff1997} which is isostructural to PbCuTe$_2$O$_6$.\cite{Koteswararao2014} Both Pb and Sr compounds share the same $P4_132$ crystallographic symmetry and feature very similar lattice parameters of 12.47\,\r A and 12.49\,\r A, respectively.\cite{Wulff1997,Koteswararao2014} Given the fact that neither Sr nor Pb are directly involved in superexchange pathways, little difference between the two compounds could be envisaged. Surprisingly, we find that SrCuTe$_2$O$_6$ is largely dissimilar to its Pb analog and exhibits quasi-one-dimensional magnetism, which is rationalized microscopically by the dominant third-neighbor exchange. At low temperatures, two consecutive magnetic transitions are observed in zero field, and altogether three distinct phase are established in applied magnetic fields up to 9~T. They may be related to the frustrated nature of interchain couplings.

\begin{figure*}
\includegraphics{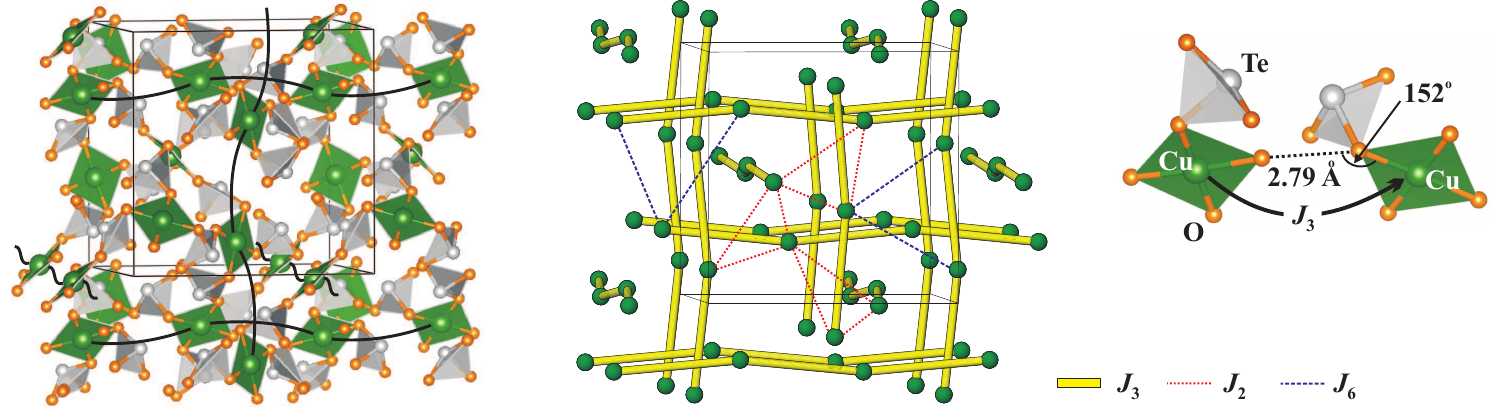}
\caption{\label{fig:structure}
Left panel: crystal structure of SrCuTe$_2$O$_6$, with solid lines denoting spin chains. The Sr atoms are not shown. Middle panel: spin lattice of SrCuTe$_2$O$_6$ composed of crossing spin chains. The interchain couplings are frustrated for both $J_2$ and $J_6$ shown by dotted and short-dashed lines, respectively. Right panel: Cu--O$\ldots$O--Cu superexchange pathway for the intrachain coupling $J_3$. Note that this coupling does not involve TeO$_3$ pyramids.
}
\end{figure*}

The common crystal structure of SrCuTe$_2$O$_6$ and PbCuTe$_2$O$_6$ is shown in Fig.~\ref{fig:structure}. It is formed by CuO$_4$ plaquettes and TeO$_3$ trigonal pyramids that can be alternatively viewed as TeO$_3$E pyramids with the lone pair E occupying one of the vertices. The lone pair originates from the $5s^2$ electronic configuration of Te$^{4+}$ and triggers a pronounced asymmetry of the Te coordination environment. The crystal structure is non-centrosymmetric, and the magnetic Cu$^{2+}$ ions take the $12d$ Wyckoff position on the two-fold rotation axis passing along face diagonal of the cubic unit cell.

\section{Methods}
Synthesis of SrCuTe$_{2}$O$_{6}$ was purely accidental. In an attempt to synthesize the polycrystalline sample of SrCuTe$_{2}$O$_{7}$ by the solid-state reaction method,\cite{Yeon2011} SrCO$_{3}$ (Aldrich, 99.995\%), CuO (Aldrich, 99.9999\%), TeO$_{2}$ (Aldrich, 99.9995\%), and H$_{2}$TeO$_{4}\cdot2$H$_{2}$O (Alfa Aesar, 99\%) were taken as initial reactants. These initial reactants were ground thoroughly in stoichiometric ratios and fired at 630~$^{\circ}$C for three days in flowing argon atmosphere with two intermediate grindings. The x-ray powder diffraction (XRD) pattern was recorded at room temperature using PANalytical powder diffractometer (CuK$_{\alpha}$ radiation, $\lambda_{\rm avg} = 1.54182$~\AA). The final product was phase-pure SrCuTe$_{2}$O$_{6}$ instead of SrCuTe$_{2}$O$_{7}$. The loss of oxygen is, presumably, related to reducing synthesis conditions, whereas annealings in air may be required to synthesize SrCuTe$_{2}$O$_{7}$.

\begin{figure}
\includegraphics{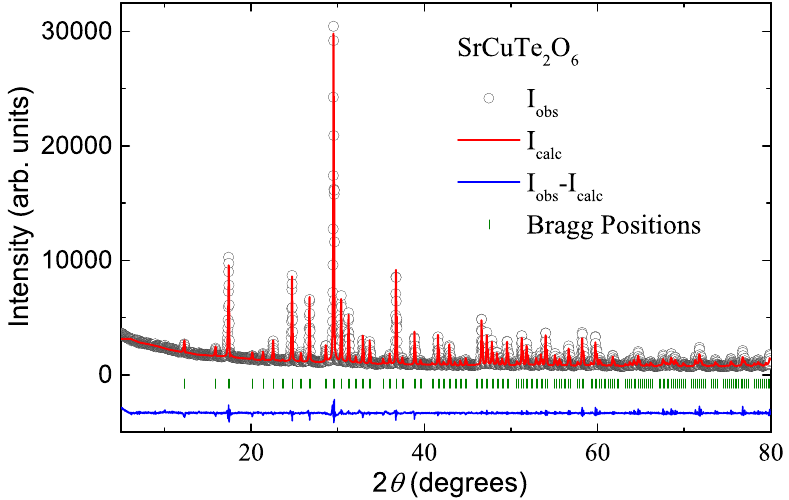}
\caption{\label{Fig2} Powder x-ray diffraction data of SrCuTe$_{2}$O$_{6}$ collected at room temperature. The solid line represents the Le Bail fit of the data. The Bragg peak positions are indicated by green vertical bars and the bottom solid blue line indicates the difference between the experimental and calculated intensities.}
\end{figure}

Le Bail fit of the powder XRD data was performed using \verb"FullProf" software package.\cite{Rodriguez1993} All peaks are consistent with the cubic structure and space group $P$4$_{1}$32 (No.~213). Figure~\ref{Fig2} displays the powder XRD data along with the fit. The initial structural parameters were taken from Ref.~\onlinecite{Wulff1997}. The best fit was obtained with a goodness-of-fit, $\chi^{2} \simeq 3.74$. The refined lattice parameters are $a=b=c=12.4681(1)$~{\AA}, which are consistent with the previous report.\cite{Wulff1997}

Magnetization ($M$) measurements were performed as a function of temperature $T$ and magnetic field $H$ using vibrating sample magnetometer (VSM) attachment to the Physical Property Measurement System (PPMS, Quantum Design). Heat capacity $C_{\rm p}$ data were measured with a heat capacity attachment to PPMS as a function of $T$ and $H$ on a sintered pellet using the relaxation technique.

Density-functional (DFT) band structure calculations were performed using the \texttt{FPLO} code\cite{fplo} and generalized gradient approximation (GGA)\cite{pbe96} for the exchange-correlation potential. Strong correlation effects in the Cu $3d$ shell were taken into account by including a mean-field GGA+$U$ correction with the Hubbard repulsion parameter $U_d=8.5$\,eV and Hund's coupling $J_d=1$\,eV.\cite{Mazurenko2014,Nath2014a} All calculations were performed for the crystallographic unit cell with 120 atoms. Reciprocal space was sampled by 20 \textbf{k}-points in the symmetry-irreducible part of the first Brillouin zone. This sampling is already sufficient for a decent convergence of total energies and ensuing magnetic couplings, given the huge size of the unit cell in the direct space.

\section{Results}

\subsection{Magnetization}
\begin{figure}
\includegraphics{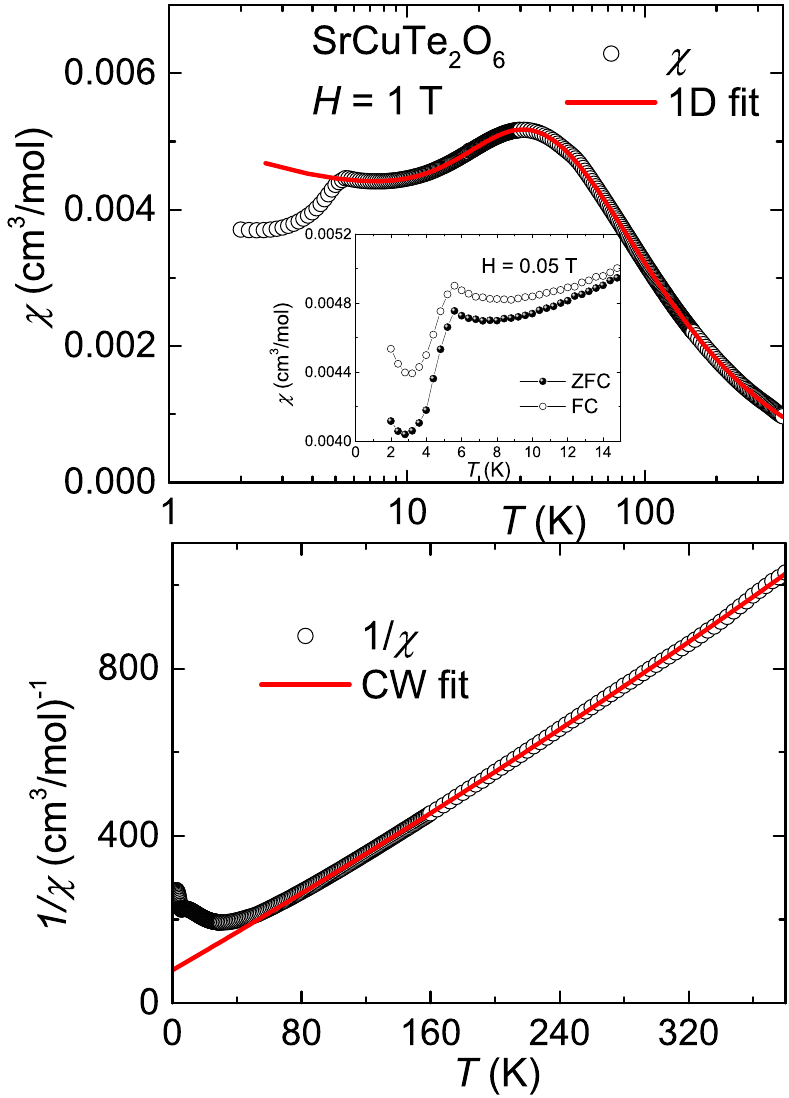}
\caption{\label{Fig3} Upper panel: Magnetic susceptibility ($\chi$) of SrCuTe$_{2}$O$_{6}$ as a function of temperature measured at an applied magnetic field of 1~T. The solid red line represents the 1D model fit with Eq.~\eqref{Johnston} as described in the text. The inset shows the ZFC and FC susceptibilities measured at $H=0.05$~T in the low-temperature regime. Lower panel: Inverse susceptibility $1/\chi$ as a function of temperature. The solid line is the CW fit with Eq.~\eqref{cw}.}
\end{figure}
The magnetic susceptibility $\chi$ ($= M/H$) measured in the temperature range $2~{\rm K} \leq T \leq 380$~K and in an applied magnetic field ($H$) of 1 T is shown in the upper panel of Fig.~\ref{Fig3}. It follows a Curie-Weiss (CW) behavior at high temperatures, as expected in the paramagnetic regime, and passes through a broad maximum at around $T_{\chi}^{\rm max} \simeq 30$~K. This broad maximum is a hallmark of short-range-order anticipated for low-dimensional (AFM) spin systems. Its position in temperature is a measure of the exchange energy.\cite{Bonner1964,Eggert1994,Johnston2000} With decreasing temperature, a sharp peak is observed at $T_{\rm N1} \simeq 5.5$~K suggesting a transition toward long-range magnetic order (LRO). At very low temperatures, $\chi(T)$ is increasing slightly, which is likely due to some extrinsic paramagnetic impurities or defect/dislocations present in the powder sample.

The high-temperature data were analyzed by fitting $\chi(T)$ with the following expression:
\begin{equation}
\chi(T)=\chi_0+\frac{C}{T+\theta_{\rm CW}},
\label{cw}
\end{equation}
where $\chi_0$ is the temperature-independent susceptibility including contributions of core diamagnetism and Van-Vleck paramagnetism, and the second term is the CW law with $C$ being the Curie constant and $\theta_{\rm CW}$ the characteristic Curie-Weiss temperature.

Our fit above 150~K (lower panel of Fig.~\ref{Fig3}) yields $\chi_0 \simeq -1.159(5) \times 10^{-4}$~cm$^{3}$/mol, $C \simeq 0.45(2)$~cm$^{3}$\,K/mol, and $\theta_{\rm CW} \simeq 35.4(1)$~K. From the value of $C$, the effective moment ($\mu_{\rm eff} = \sqrt{3k_{\rm B}C/N_{\rm A}}$) was calculated to be $\simeq 1.9~\mu_{\rm B}$ where $N_{\rm A}$ is Avogadro's number and $k_{\rm B}$ is the Boltzmann constant. This effective moment is comparable to the spin-only value of $\mu_{\rm eff} = g\sqrt{S(S+1)}\mu_{\rm B} \simeq 1.73~\mu_{\rm B}$ for Cu$^{2+}$ ($S=\frac12$) assuming $g=2$. Given the fact that the experimental moment is above the spin-only value, we anticipate $g\simeq 2.20$, which is in the typical range for Cu$^{2+}$ compounds.\cite{Nath2014a} The core diamagnetic susceptibility ($\chi_{\rm core}$) of SrCuTe$_{2}$O$_{6}$ was calculated to be $-1.52\times 10^{-4}$~cm$^{3}$/mol by adding the $\chi_{\rm core}$ of the individual ions Sr$^{2+}$, Cu$^{2+}$, Te$^{4+}$, and O$^{2-}$.\cite{Selwood1956} The Van-Vleck paramagnetic susceptibility ($\chi_{\rm VV}$) was calculated by subtracting $\chi_{\rm core}$ from $\chi_0$ to be $3.6(2)\times 10^{-5}$~cm$^{3}$/mol.
This value of $\chi_{\rm VV}$ is comparable with other cuprate compounds, like Sr$_2$CuO$_3$,\cite{Motoyama1996} Sr$_2$CuP$_2$O$_8$,\cite{Nath2005} PbCuTe$_{2}$O$_{6}$,\cite{Koteswararao2014} and PbCu$_{3}$TeO$_{7}$.\cite{Koteswararao2013}
The positive value of $\theta_{\rm CW}$ suggests that the dominant exchange interactions between Cu$^{2+}$ ions are AFM in nature.

As shown in Fig.~\ref{Fig3}, $\chi(T)$ reveals a pronounced maximum at $T_{\chi}^{\rm max} \simeq 30$~K suggesting that the spins in SrCuTe$_{2}$O$_{6}$ form a short-range order before entering the LRO state. Such a susceptibility maximum is expected for a variety of (mostly low-dimensional) spin models. Here, we refer to the simplest possible case of a uniform spin chain. Apart from providing an excellent fit of the experimental data, the chain model is well justified by the microscopic analysis presented in Sec.~\ref{sec:microscopic} below.

To fit the bulk $\chi(T)$ data, we decomposed $\chi$ into three components
\begin{equation}
 \chi(T)=\chi_0+\frac{C_{\rm imp}}{T+\theta_{\rm imp}} + \chi_{\rm 1D}(T) ,
 \label{Johnston}
\end{equation}
where the second term is the CW law that takes into account the impurity contribution. $C_{\rm imp}$ gives information about the impurity concentration, $\theta_{\rm imp}$ provides an effective interaction between impurity spins, and $\chi$$_{\rm 1D}(T)$ is the expression for spin susceptibility of uniform one-dimensional (1D) Heisenberg spin-$\frac12$ AFM chain given by Johnston $\textit{et al}$.\cite{Johnston2000} This expression is valid over a wide temperature range. The fit of $\chi(T)$ data above 7~K ($T > T_{N1}$) by Eq.~\eqref{Johnston} is shown in the upper panel of Fig.~\ref{Fig3}. The best fit down to 7~K, where the LRO transition is approached, was obtained with the parameters $\chi_0 = -1.192(2) \times 10^{-4}$~cm$^{3}$/mol, $C_{\rm imp}$ = 0.0034(1)~cm$^{3}$K/mol, $\theta$$_{\rm imp}$ = 1.49(3)~K, $g$ = 2.156(1), and $J/k_{\rm B}$ = 49.34(1)~K. The low value of $\theta_{\rm imp}$ suggests a weak interaction among the impurity spins. Similarly, the obtained value of $C_{\rm imp}$ corresponds to of 0.74\% impurity spins assuming that they have spin-$\frac12$ nature.
\begin{figure}
\includegraphics{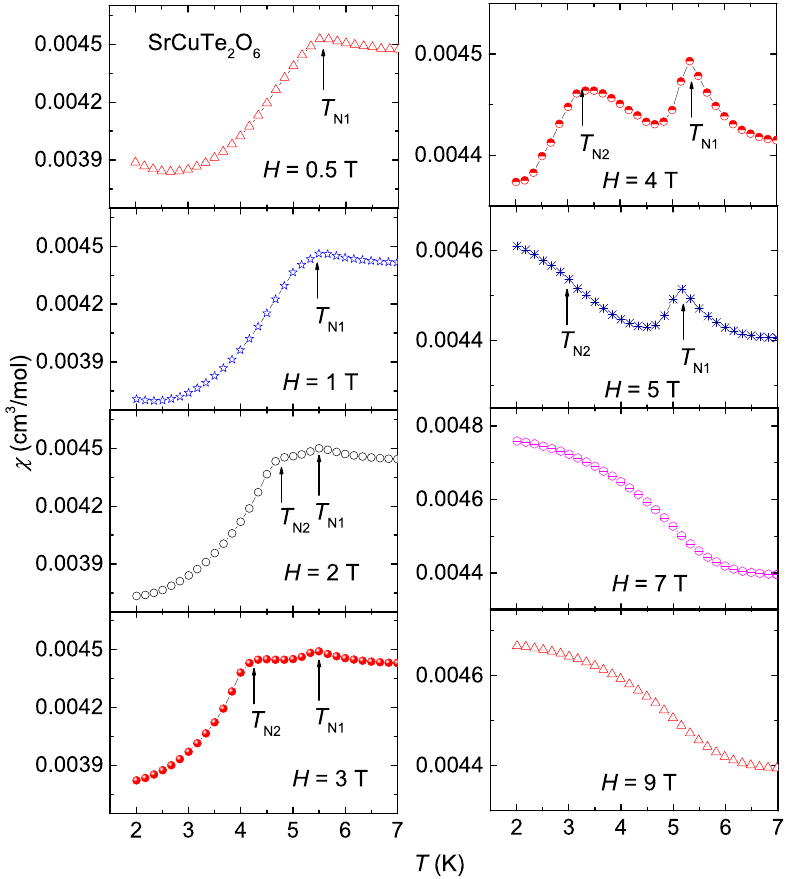}
\caption{\label{Fig4} Low-temperature $\chi(T)$ of SrCuTe$_{2}$O$_{6}$ measured at different applied fields between 0.5~T and 9~T. The peaks associated with two transitions ($T_{\rm N1}$ and $T_{\rm N2}$) are shown by upward arrows.}
\end{figure}

Zero-field cooled (ZFC) and field-cooled (FC) susceptibilities as a function of temperature measured in the magnetic field of 0.05~T are shown in the inset of Fig.~\ref{Fig3}. Both curves showed a sharp peak at 5.5~K without any significant splitting divergence, thus ruling out the possibility of a spin-freezing/spin-glass transition at low temperatures.

\begin{figure}
\includegraphics{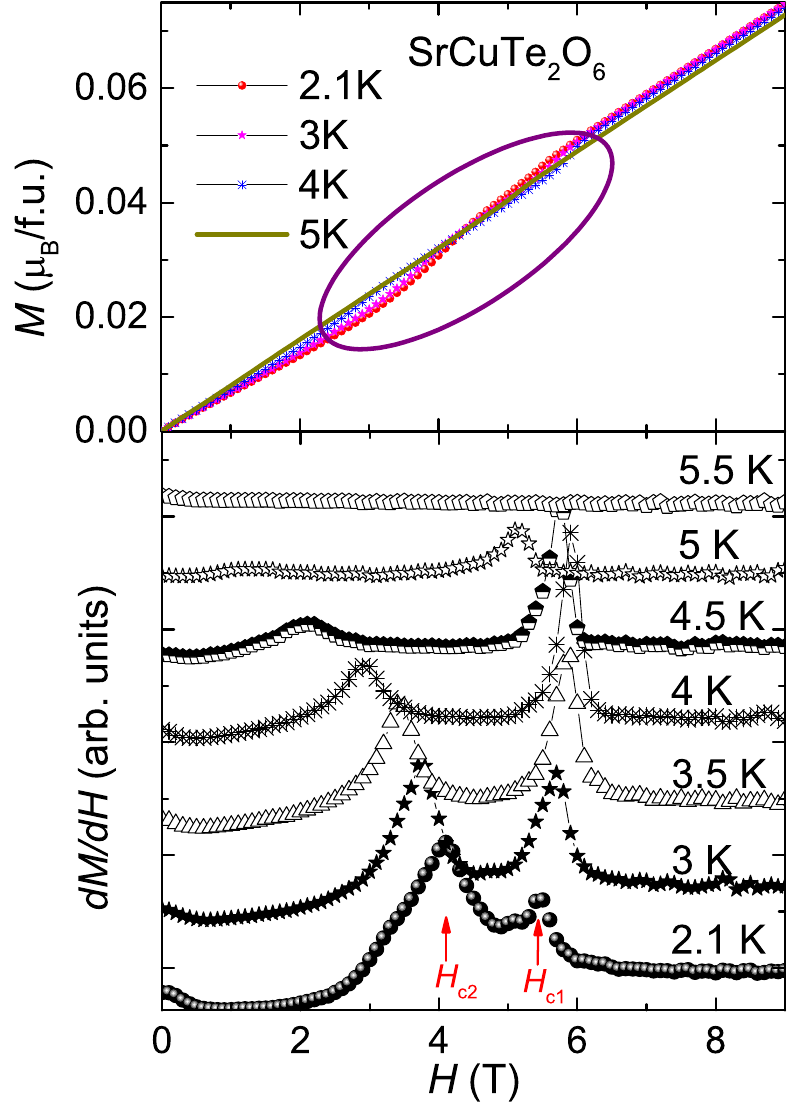}
\caption{\label{Fig5} Upper panel: Magnetization vs. applied field for SrCuTe$_{2}$O$_{6}$ measured at various temperatures from 2.1~K to 5~K. The encircled region shows the field-induced transition. Lower panel: The derivative of $M$ with respect to $H$ ($dM/dH$) at various temperatures from 2.1~K to 5.5~K are plotted in order to highlight two transitions. For clarity, the data sets at various temperatures are added by an offset along the $y$-axis.}
\end{figure}
In order to probe the magnetic transition, $\chi(T)$ was measured at different applied fields in the low-temperature regime (Fig.~\ref{Fig4}). At $H$ = 0.5~T, it shows a clear peak at $T_{\rm N1} \simeq$ 5.5~K. With increasing $H$, the peak was found to broaden. At $H = 2$~T, this broad peak transforms into two peaks at $T_{\rm N1} \simeq$ 5.5~K and $T_{\rm N2} \simeq$ 4.7~K. With further increase in $H$, the position of $T_{N1}$ remains almost unchanged, while the feature at $T_{N2}$ moves to lower temperatures and broadens substantially. At $H$ = 5~T, the trend below $T_{N2}$ changes, and the susceptibility increases toward low temperatures, as in the spin-flop phase of more conventional Cu$^{2+}$ antiferromagnets.\cite{Janson2011} For $H >$ 6 T, the feature at $T_{\rm N2}$ disappears completely, whereas the feature at $T_{N1}$ turns into a bend of the susceptibility curve that is reminiscent of the magnetic transition in canted antiferromagnets. We also tried to detect magnetic transitions by plotting $d\chi/dT$, but no extra features could be seen there.

In order to gain additional insight into the nature of magnetic transitions observed in $\chi(T)$, we measured magnetization isotherms $M(H)$ at different temperatures. In polycrystalline samples, spin-flop transition manifests itself by kinks of magnetization curves. Indeed, in our case, we observed such kinks in the magnetization isotherms measured at $T \leq$ 5~K (Fig.~\ref{Fig5}, upper panel). Transition fields are determined from field derivatives of the magnetization curves plotted in the lower panel of Fig.~\ref{Fig5}. At $T$ = 2.1~K, two clear peaks are observed at the critical fields marked as $H_{\rm c1}$ and $H_{\rm c2}$. As we increase the temperature, $H_{\rm c2}$ moves rapidly towards lower magnetic fields, whereas $H_{\rm c1}$ increases slightly till $T = 4$~K, where it starts decreasing and then disappears completely for $T>5$~K, the temperature that nearly coincides with $T_{N1}$. On the other hand, the temperature evolution of $H_{\rm c2}$ is clearly reminiscent of the field evolution of $T_{N2}$ (see also Fig.~\ref{Fig4}).

\begin{figure}
\includegraphics{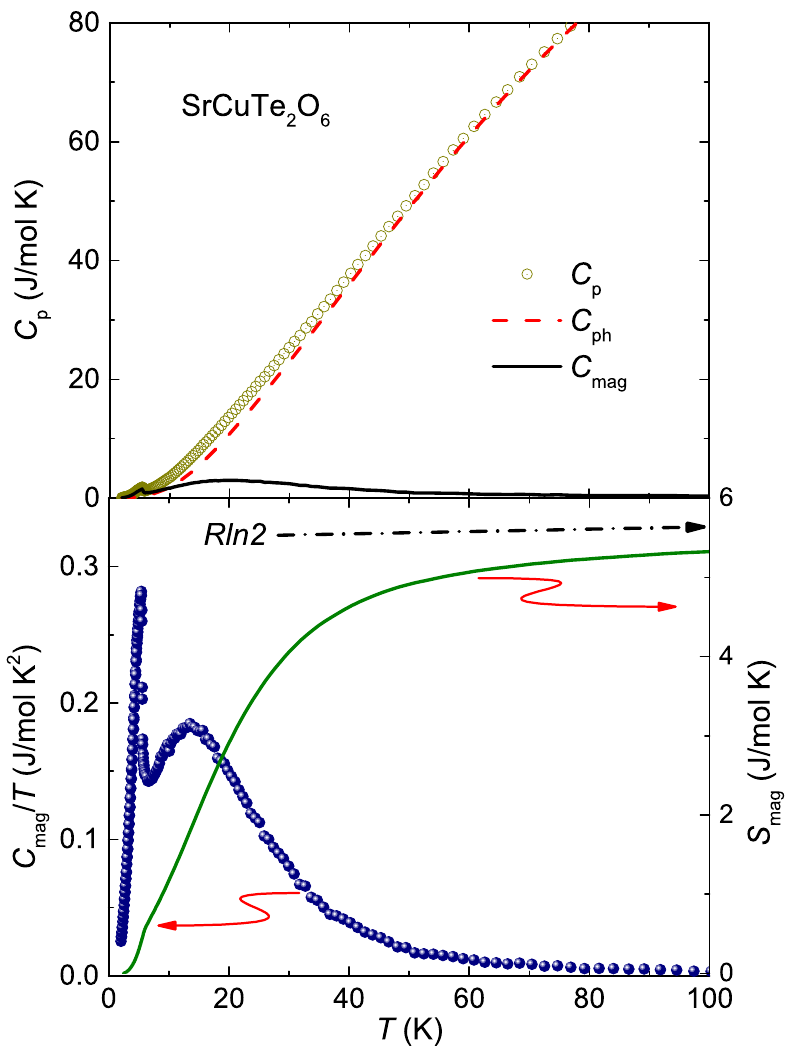}
\caption{\label{Fig6} Upper panel: Heat capacity $C_{\rm p}$ vs. $T$ of SrCuTe$_{2}$O$_{6}$ in zero applied magnetic field. The dashed red line is the phonon contribution to the specific heat $C_{\rm ph}$ using Debye fit (Eq.~\ref{Debye}). The black solid line indicates the magnetic contribution to the heat capacity $C_{\rm mag}$. Lower panel: The left $y$-axis shows the ($C_{\rm mag}/T$), and the right $y$-axis shows the magnetic entropy ($S_{\rm mag}$) versus temperature $T$.}
\end{figure}
\subsection{Heat capacity}
\label{sec:heat}
In magnetic insulators, $C_{\rm p}$ has two principal contributions: phononic part and magnetic part.
Figure.~\ref{Fig6} shows the heat capacity $C_{\rm p}$ as a function of temperature measured in zero field. At high temperatures, it is completely dominated by the phonon contributions and then shows a lambda-type anomaly at $T_{\rm N1} \simeq$ 5.4~K followed by a bend in the vicinity of $T_{N2}$. To obtain the phononic part of the heat capacity ($C_{\rm ph}$), the raw data at high temperatures were fitted by a linear combination of five Debye functions:\cite{Nath2008,*Nath2008a}
\begin{equation}
\label{Debye}
C_{\rm ph}(T) = 9R\times\sum\limits_{\rm n=1}^{5} c_{\rm n} \left(\frac{T}{\theta_{\rm Dn}}\right)^3 \int_0^{\frac{\theta_{\rm Dn}}{T}} \frac{x^4e^x}{(e^x-1)^2} dx.
\end{equation}
Here, $R$ is the universal gas constant, the coefficients $c_{\rm n}$ represent groups of distinct atoms in the crystal, and $\theta_{\rm Dn}$ are the corresponding Debye temperatures.
%The Debye temperatures $\theta_{\rm D1}$, $\theta_{\rm D2}$, $\theta_{\rm D3}$, $\theta_{\rm D4}$ and $\theta_{\rm D5}$ are 109~K, 259~K, 231~K, %367~K and 954~K and the ratio of associated coefficients $c_{1}$:$c_{2}$:$c_{3}$:$c_{4}$:$c_{5}$ are fixed to 1:1:1:2:5.
The magnetic part of the heat capacity ($C_{\rm mag}$) was obtained by subtracting $C_{\rm ph}$ from the total heat capacity $C_{\rm p}$. The subtraction procedure was verified by calculating the magnetic entropy $S_{\rm mag}$ through the integration of $C_{\rm mag}(T)$/$T$ that yields $S_{\rm mag} \simeq$ 5.4~J\,mol$^{-1}$\,K$^{-1}$ at 150~K (lower panel of Fig.~\ref{Fig6}). This value is not far from the expected full magnetic entropy for spin-$\frac12$: $S_{\rm mag} = R\ln 2$ = 5.76 J\,mol$^{-1}$\,K$^{-1}$.
\begin{figure}
\includegraphics{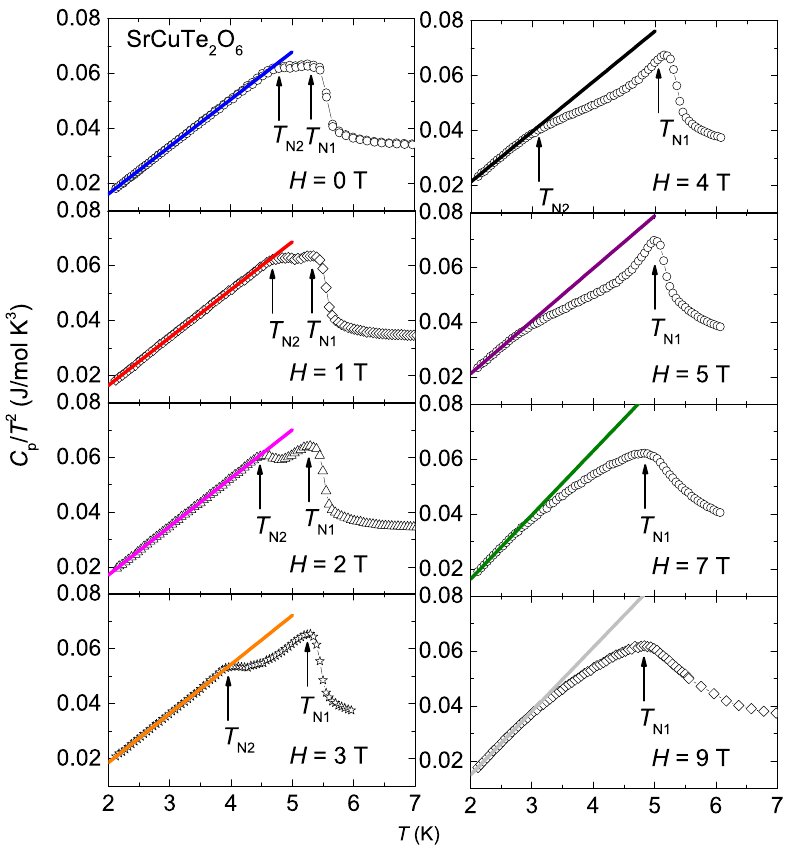}
\caption{\label{Fig7} $C_{\rm p}/T^{2}$ vs. $T$ measured under different magnetic fields from 0 to 9~T. The transitions at $T_{\rm N1}$ and $T_{\rm N2}$ are indicated by upward arrows. The solid line shows a fit of ($C_{\rm p} \propto T^{3}$) in the low-temperature regime.}
\end{figure}

Figure~\ref{Fig7} shows the plot of $C_{\rm p}$/$T^{2}$ as a function of $T$ measured at different applied fields from 0 to 9~T. At $H$ = 0~T, two transitions are clearly visible at $T_{\rm N1}$ and $T_{\rm N2}$. With increase in $H$, $T_{\rm N1}$ shows a very weak effect, while $T_{\rm N2}$ moves towards low temperatures. For $H >$ 4~T, the peak associated with $T_{\rm N2}$ broadens and eventually disappears. At $T \leq T_{\rm N2}$, $C_{\rm p}$ follows a $T^{3}$ behavior that indicates 3D spin-wave dispersions in the ordered state.\cite{Nath2014}
\begin{figure}
\includegraphics{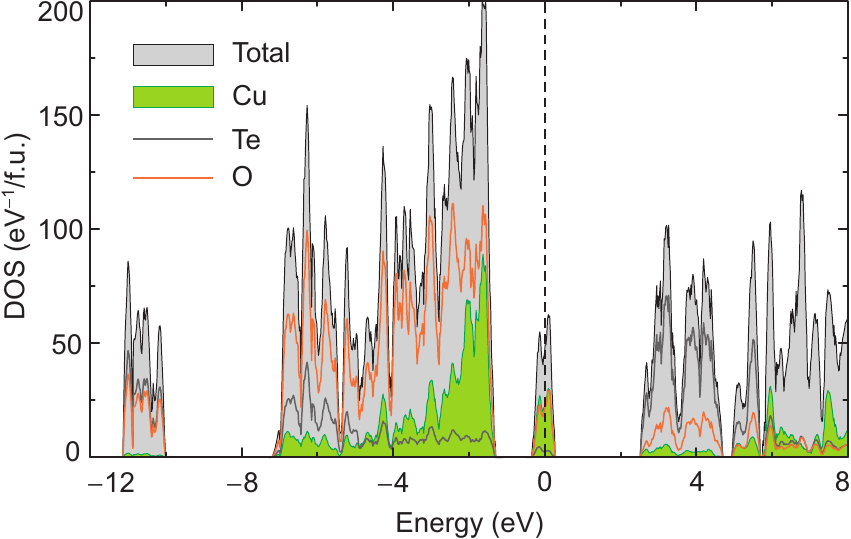}
\caption{\label{Fig8} GGA density of states (DOS) for SrCuTe$_2$O$_6$. The Fermi level is at zero energy.}
\end{figure}

\subsection{Microscopic magnetic model}
\label{sec:microscopic}
The quasi-1D behavior of SrCuTe$_2$O$_6$ is, at first glance, perplexing, given the 3D nature of the crystal structure (Fig.~\ref{fig:structure}). In order to rationalize this behavior, we evaluated individual exchange couplings in SrCuTe$_2$O$_6$. Two complementary procedures, the model analysis and total-energy calculations, were used. The model analysis rests upon a tight-binding fit of the GGA band structure (Fig.~\ref{Fig8}). In GGA, strong correlation effects are largely neglected and should be, hence, taken into account on a model level by supplying the GGA-based tight-binding Hamiltonian with the Hubbard term, where $U_{\rm eff}=4.5$~eV\cite{Janson2011,Rousochatzakis2015} stands for an effective Coulomb repulsion in the Cu--O bands. This way, AFM parts of the exchange couplings are calculated as $J_i^{\rm AFM}=4t_i^2/U_{\rm eff}$, where $t_i$ are tight-binding (hopping) parameters.

\begin{table}
\caption{\label{tab:hoppings}
Exchange couplings in SrCuTe$_2$O$_6$: Cu--Cu distances $d$ (in\,\r A), hopping parameters of the tight-binding model $t_i$ (in\,meV), AFM contributions to the exchange couplings $J_i^{\rm AFM}=4t_i^2/U_{\rm eff}$ (in\,K), and total exchange couplings $J_i$ (also in\,K) obtained from GGA+$U$ calculations, as described in the text. The couplings not listed in this Table are well below 1\,K.
}
\begin{ruledtabular}
\begin{tabular}{ccrrr}
  & $d_{\text{Cu--Cu}}$ & $t_i$ & $J_i^{\rm AFM}$ & $J_i$ \\
  $J_1$ & 4.55 & 11    &  1 & 0.3 \\
  $J_2$ & 5.52 & 27    &  8 & 4   \\
  $J_3$ & 6.29 & $-83$ & 71 & 45  \\
  $J_6$ & 8.96 & $-20$ &  4 & 2   \\
\end{tabular}
\end{ruledtabular}
\end{table}

\begin{figure}
\includegraphics{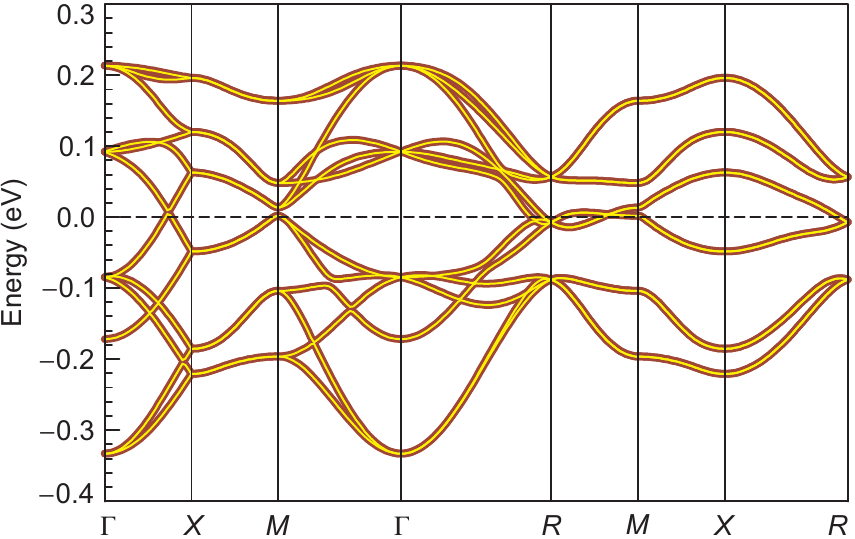}
\caption{\label{Fig9} GGA Cu $d_{x^2-y^2}$ bands in the vicinity of the Fermi level (thin lights lines) and their fit with the tight-binding model (thick dark lines). }
\end{figure}

Hopping parameters were calculated by constructing Wannier functions for the isolated 12-band complex at the Fermi level (Fig.~\ref{Fig9}). These 12 bands originate from $d_{x^2-y^2}$ states of 12 Cu atoms in the unit cell. Here, the $x$ and $y$ axes are directed along the Cu--O bonds, and the magnetic $x^2-y^2$ orbital lies in the CuO$_4$ plane, similar to the majority of cuprates. The hopping parameters listed in Table~\ref{tab:hoppings} reveal the leading third-neighbor AFM coupling $J_3$, whereas the couplings between first, second, and sixth neighbors are about order of magnitude smaller. Other couplings, including those between fourth and fifth neighbors, yield $t_i<5$\,meV, hence $J_i\ll 1$\,K.

The results of the model analysis were verified by a direct calculation of exchange integrals $J_i$ from total energies of collinear spin configurations obtained in GGA+$U$, where correlation effects are directly included in the self-consistent DFT procedure.\footnote{This calculation resulted in a band gap of about 3.0\,eV for the lowest-energy AFM spin configuration.} These exchange integrals are in excellent agreement with those obtained from the model analysis. The dominant interaction is $J_3\simeq 45$\,K in good agreement with the experimental $J_{\rm 1D}\simeq 49.3$\,K. The leading interchain couplings are $J_2$ and $J_6$.

Let us now consider the topology of the spin lattice. According to Ref.~\onlinecite{Koteswararao2014}, $J_1$ forms isolated triangles that are basically irrelevant to the physics of SrCuTe$_2$O$_6$ because $J_1$ is nearly zero. The second-neighbor coupling $J_2$ forms a hyperkagome network (Fig.~\ref{fig:structure}, middle). The leading coupling $J_3$ forms perpendicular chains running along the $a$, $b$, and $c$ directions (Fig.~\ref{fig:structure}). Finally, $J_6$ forms another complex network that also frustrates the lattice, because each atom of a given $J_3$ chain is coupled to two neighboring atoms of another chain (Fig.~\ref{fig:structure}, middle). We thus find that SrCuTe$_2$O$_6$ is a spin-chain compound with a complex and strongly frustrated network of interchain couplings.

Remarkably, none of the exchange couplings in SrCuTe$_2$O$_6$ entails a direction connection between the CuO$_4$ plaquettes, and even those plaquettes that are connected via TeO$_3$ pyramids reveal a weak coupling $J_2\simeq 4$\,K only. The leading interaction runs between third neighbors despite the fact that the respective CuO$_4$ plaquettes lack any obvious structural connectivity (Fig.~\ref{fig:structure}, right). Nevertheless, these plaquettes are arranged in such a way that two Cu--O bonds are directed toward each other, the O$\ldots$O distance is about 2.79\,\r A, and the resulting Cu--O$\ldots$O--Cu pathway is only weakly bent with the Cu--O$\ldots$O angle of 152$^{\circ}$ compared to 180$^{\circ}$ for the linear pathway. This configuration triggers a sizable exchange coupling of nearly 50\,K for the Cu atoms that are more than 6\,\r A apart, similar to other cuprate materials, where individual CuO$_4$ units are structurally disconnected. By contrast, shorter Cu--Cu distances observed for $J_1$ and $J_2$ feature strongly bent Cu--O$\ldots$O--Cu pathways and turn out to be inefficient for the superexchange.

\section{Discussion and summary}
\begin{figure}
\includegraphics{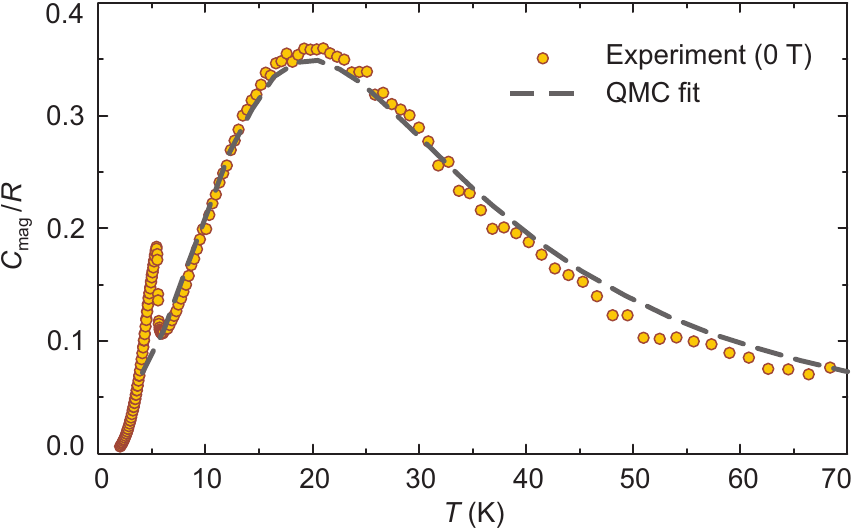}
\caption{\label{Fig10} $C_{\rm mag}/R$ vs. $T$ of SrCuTe$_{2}$O$_{6}$ where $R = 8.314$~J/mol~K is the universal gas constant. The solid line is the quantum Monte-Carlo (QMC) simulation\cite{Johnston2000} for the uniform Heisenberg spin-$\frac12$ chain with $J_{\rm 1D}=42$~K.}
\end{figure}

%The one-dimensional Heisenberg chain model by Bonner and Fisher\cite{Bonner1964,Eggert1994,Kim1998} estimates the exchange interaction from magnetic susceptibility. It is given as $T_{\rm max}^{\chi}$ $\simeq$ 0.641 $J_{\rm 1D}$. From $T_{\rm max}^{\chi}$ $\simeq$ 30~K, we get $J_{\rm 1D}$ $\simeq$ 46.8147~K. For a two dimensional system, if we consider the antiferromagnetic square lattice model, the value of $J_{\rm 2D}$ estimated from the relation $T_{\rm max}^{\chi}$ $\simeq$ 0.94 $J_{\rm 2D}$\cite{Kim1998} is $J_{\rm 2D}$ $\simeq$ 31.9148~K. The DFT calculations agree that the dominant exchange interaction is of chain type with interaction strength $J_{\rm 1D}$ $\simeq$ 45~K.
When probed above $T_{\rm N1}$, SrCuTe$_2$O$_6$ is, at first glance, indistinguishable from a conventional spin-$\frac12$ Heisenberg chain antiferromagnet. It is a quantum spin system showing broad maxima in the magnetic susceptibility and magnetic specific heat. The positions of these maxima and the absolute values at the maxima are, in general, consistent with the quantum spin-chain model. The maximum value of $C_{\rm mag}$ ($C_{\rm mag}^{\rm max}/R \simeq$~0.359) is higher than in typical two-dimensional and three-dimensional frustrated spin systems and lower than in a non-frustrated two-dimensional quantum antiferromagnet,\cite{Bernu2001,Hofmann2003} thus indicating the 1D nature of the spin lattice. The curve above $T_{\rm N1}$ follows the specific heat of a uniform quantum spin-$\frac12$ chain (Fig.~\ref{Fig10}),\cite{Johnston2000} although we had to re-normalize the exchange coupling to $J_{\rm 1D}\simeq 42$~K compared to $J_{\rm 1D}\simeq 49$~K from the susceptibility fit. The origin of this discrepancy is not entirely clear. The shift of the specific heat maximum toward lower temperatures may be due to enhanced quantum fluctuations arising from frustrated nature of the interchain couplings. However, we cannot completely exclude artifacts related to the data analysis and to the subtraction of the phonon contribution.

\begin{figure}
\includegraphics [width=3in]{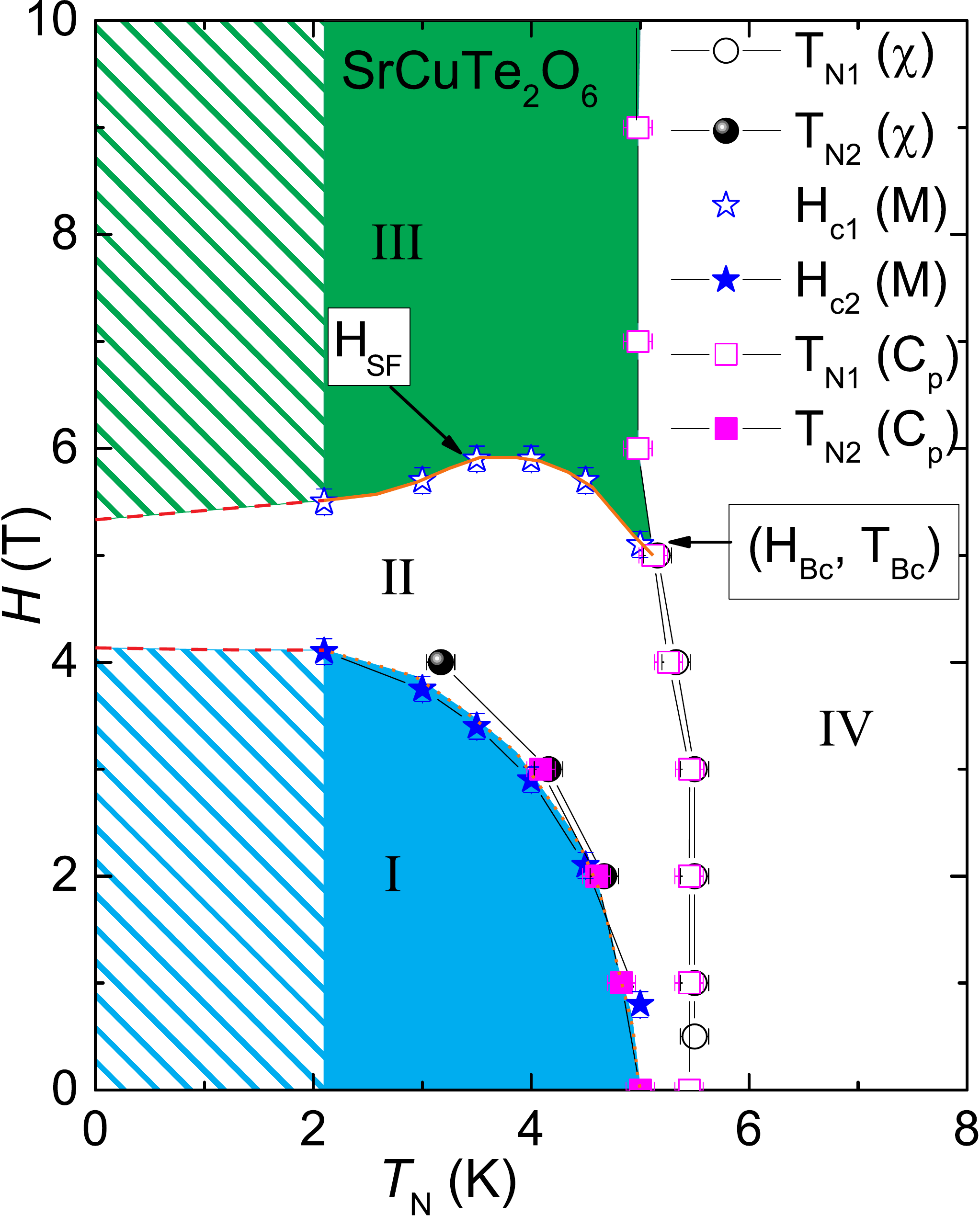}
\caption{\label{Fig11} $H-T$ phase diagram of SrCuTe$_{2}$O$_{6}$ by picking $T_{\rm N1}$ and $T_{\rm N2}$ from both $\chi$ and $C_{\rm p}$ measurements and $H_{\rm c1}$ and $H_{\rm c2}$ from the $M(H)$ measurements. The SF phase transition line (dashed line) is drawn using $H_{\rm c1}$ values obtained from the $M(H)$ measurements. The point($H_{\rm Bc}$, $T_{\rm Bc}$) at which the SF transition ends is called bi-critical point. Error bars are generally comparable to the symbol size. Dashed lines and hatched fillings below 2~K indicate the tentative nature of the phase diagram in this temperature range, as we have not done measurements below 2~K.}
\end{figure}

At low temperatures, the majority of spin-chain systems undergo a single magnetic transition toward the long-range-ordered antiferromagnetic state. Disregarding the geometry of interchain couplings, transition temperature can be determined from the expression proposed by Schulz:\cite{Schulz1996}
\begin{equation}
 |J_{\perp}| \simeq \frac{T_{\rm N}}{1.28\sqrt{\ln(5.8J/T_{\rm N})}},
 \label{Schulz}
\end{equation}
where $J_{\perp}$ is an effective interchain coupling, and four interchain couplings per Cu$^{2+}$ ion are assumed. Taking $T_{\rm N1}\simeq$~5.5~K and $J/k_{\rm B} \simeq$~49.3~K, we arrive at $|J_{\perp}| \simeq$~2.2~K, which is, remarkably, of the same magnitude as the interchain couplings $J_2$ and $J_6$ derived from our microscopic analysis (Table~\ref{tab:hoppings}). A more advanced approach of Ref.~\onlinecite{Yasuda2005} yields a very similar value of $|J_{\perp}|\simeq$~2.5~K. Therefore, we expect that the frustration of interchain couplings in SrCuTe$_2$O$_6$ has little influence on the value of $T_{\rm N}$, but it may be responsible for the presence of two transitions and for the unusual temperature-field phase diagram (Fig.~\ref{Fig11}).

Our data suggest that the magnetic order sets in below $T_{\rm N1}$. The $\lambda$-type anomaly of the specific heat indicates a conventional second-order transition. The second transition at $T_{\rm N2}$ (also tracked by $H_{c2}$) manifests itself by a hump in the specific heat and could be a spin-reorientation transition. Yet another transition takes place in the magnetic field of $5-6$\,T at $H_{c1}$. The weak field dependence of the respective line on the phase diagram (Fig.~\ref{Fig11}) is reminiscent of a spin-flop transition in conventional antiferromagnets. Similar to the spin-flop transition, this line ends at a bicritical point where the magnetic ordering transition at $T_{\rm N1}$ is reached. Altogether, our data identify three distinct ordered phases of SrCuTe$_2$O$_6$ (I, II, and III) on the $H\!-\!T$ phase diagram up to 9\,T (Fig.~\ref{Fig11}). Additional phases could be present at higher fields. From $J_{\rm 1D}\simeq 49$\,K, the saturation field as high as $H_s=2J_{\rm 1D}(k_B/g\mu_B)\simeq 68$\,T should be expected.

Although SrCuTe$_2$O$_6$ is a quasi-1D spin-chain compound, its $H\!-\!T$ phase diagram is reminiscent of the behavior of triangular-lattice antiferromagnets,\cite{Seabra2011,Gvozdikova2011} where the $120^{\circ}$ antiferromagnetic order, which is the ground state of the Heisenberg antiferromagnet on the triangular lattice,\cite{Capriotti1999,*White2007} competes with the so-called up-up-down ($uud$) phase. The $uud$ phase is stabilized by the magnetic field that introduces axial (Ising) anisotropy and disfavors the non-collinear $120^{\circ}$ order. Thermal fluctuations may also stabilize the collinear $uud$ phase in zero field. This results in two zero-field transitions, with the $uud$ phase formed below $T_{\rm N1}$ and the $120^{\circ}$ order formed below $T_{\rm N2}$, as in Ba$_3$(Mn,Co)Nb$_2$O$_9$.\cite{Lee2014a,Lee2014b} Other triangular antiferromagnets reveal the $uud$ phase in applied magnetic fields only.\cite{[{For example: }][{}]Smirnov2007,*Hwang2012} At higher fields, yet another, 0-coplanar or canted $uud$ phase, is stabilized.\cite{Seabra2011,Gvozdikova2011,Yamamoto2015} The transition between the $uud$ and 0-coplanar phases is typically weakly field-dependent, so it can be paralleled to the II--III transition line in Fig.~\ref{Fig11}.

Whilst being a useful reference point, the scenario of multiple magnetic transitions in triangular-lattice antiferromagnets cannot be directly applied to SrCuTe$_2$O$_6$. First, the spin lattice of SrCuTe$_2$O$_6$ is not triangular. It is primarily 1D, whereas interchain couplings form an intricate network that, despite being built of triangles, can not be considered as a standard triangular lattice. Second, the physics of triangular antiferromagnetic materials is often influenced by the single-ion anisotropy that vanishes for the genuine \mbox{spin-$\frac12$} Cu$^{2+}$ ion. Therefore, other sources of anisotropy, such as Dzyaloshinsky-Moriya couplings allowed by the chiral $P4_132$ symmetry of the crystal structure, may be important. A direct investigation of the ordered phases of SrCuTe$_2$O$_6$ with neutron scattering and resonance techniques would be very interesting. To the best of our knowledge, multiple magnetic transitions in zero field and the non-trivial $H-T$ phase diagram are unique for SrCuTe$_2$O$_6$ among the broad family of quasi-1D antiferromagnets with uniform, non-frustrated spin-$\frac12$ chains. Several magnetic transitions have been observed in $\beta$-TeVO$_4$, but no information on the nature of ordered phases is presently available,\cite{Savina2011} and even the topology of spin chains has been disputed.\cite{Saul2014} LiCuVO$_4$ is another
famous example of a quasi-1D system with multiple field-induced phases,\cite{Buttgen2012,*Banks2007,schrettle2008} yet it features only one magnetic transition in zero field, whereas the complex high-field behavior may be heavily influenced by helical correlations induced by the frustration of intrachain ferromagnetic and antiferromagnetic couplings.
In summary, we have shown that SrCuTe$_2$O$_6$ is a quasi-1D chain antiferromagnet that reveals non-trivial physics at low temperatures with two ordered phases in zero field and altogether three distinct phases in the $H\!-\!T$ phase diagram studied in magnetic fields up to 9\,T. The behavior above the magnetic ordering temperature $T_{\rm N1}$ can be well understood in the framework of the Heisenberg spin-$\frac12$ chain model. The spin chains are formed by the exchange couplings between third neighbors, whereas interchain couplings are strongly frustrated. SrCuTe$_2$O$_6$ is, thus, surprisingly different from its structural sibling PbCuTe$_2$O$_6$, where first-, second- and third-neighbor couplings, all having similar magnitude, form a three-dimensional frustrated spin lattice and suppress the magnetic order below 1\,K. SrCuTe$_2$O$_6$ enters the magnetically ordered phase at a somewhat higher temperature of $T_{\rm N1}\simeq 5.5$~K in zero field and reveals an unexpectedly complex phase diagram that calls for further experimental investigation. It is worth noting that the crystal structure of SrCuTe$_2$O$_6$ is non-centrosymmetric, although non-polar. Nevertheless, ferroelectricity may be induced by the magnetic order and magnetic field, similar to LiCuVO$_4$ and other frustrated-chain magnets.\cite{schrettle2008} This possibility should be explored in future studies.

%Our $\chi(T)$ and $C_{\rm p}(T)$ data analysis along with the band structure calculations established that SrCuTe$_{2}$O$_{6}$ is an frustrated spin-1/2 Heisenberg chain compound with the intachain coupling of $J/k_{\rm B} \simeq 49$~K. The average interchain coupling was estimated to be $|J_{\perp}| \simeq$~2.16~K. A frustration parameter of $f \simeq 6.43$ primarily suggests a moderate frustration in SrCuTe$_2$O$_6$. It undergoes two successive magnetic phase transitions at low temperatures. An external applied magnetic field also leads to the formation of a SF transition below $T_{\rm N1}$. The $H-T$ phase diagram is drawn from the thermodynamic measurements which shows three distinct phases and the possible nature of the ground state in the respective phases is discussed.

\acknowledgments
NA and RN would like to acknowledge Department of Science and Technology, India for financial support. AT was funded by the ESF (Mobilitas grant MTT77), PUT733 grant of the Estonian Research Council, and by the Sofja Kovalevskaya Award of Alexander von Humboldt Foundation.

\end{document}